\documentclass{PoS}

\usepackage{graphicx}
\usepackage{amssymb}
\usepackage{amsmath}

 \newcommand{\preprint}{
   \begin{picture}(0,0)
     \put(400,130){{\rm\normalsize ADP-09-13/T691}}
   \end{picture}}

 \title{\preprint%
 Low-lying positive-parity excited states of the nucleon}

\ShortTitle{Low-lying positive-parity excited states of the nucleon}

 \author{\speaker{M. S. Mahbub} \thanks{We thank C.B. Lang for useful
     comments and discussions on our Roper results at the lattice 2009 conference.
     We thank the NCI National
     Facility and eResearch SA for  generous grants of supercomputing
     time which have enabled this project. This research is supported
     by the Australian Research Council.}\, $^{a\,b}$, Alan
   $\acute{\rm{O}}$ Cais $^{a\,c}$, Waseem Kamleh $^{a}$,
   B.G. Lasscock $^{a}$, \newline Derek B. Leinweber $^{a}$, Anthony
   G. Williams $^{a}$  \\ 
       \llap{$^a$} Special Research Centre for the Subatomic Structure
       of  Matter, Adelaide, South Australia 5005, Australia, and Department of Physics, University of Adelaide, South Australia 5005, Australia.\\
       \llap{$^b$} Department of Physics, Rajshahi University, Rajshahi 6205, Bangladesh.\\
       \llap{$^c$} Cyprus Institute, Guy Ourisson Builiding, Athalassa Campus, PO Box 27456, 1645 Nicosia, Cyprus.\\
         E-mail: \email {md.mahbub@adelaide.edu.au,
           a.ocais@cyi.ac.cy, waseem.kamleh@adelaide.edu.au, blasscoc@gmail.com, derek.leinweber@adelaide.edu.au, anthony.williams@adelaide.edu.au}}

\abstract{We present an overview of the correlation-matrix methods developed
recently by the CSSM Lattice Collaboration for the isolation of
excited states of the nucleon.  Of particular interest is the first
positive-parity excited-state of the nucleon known as the Roper
resonance. Using eigenvectors of the correlation matrix we construct parity and
eigenstate projected correlation functions which are analysed using
standardized methods. The robust nature of this approach for extracting the
eigenstate energies is presented.
We report the importance of using a variety of source and sink
smearings in achieving this. Ultimately the independence of the
eigenstate energies from the interpolator basis is demonstrated.  In
particular we consider $4\times 4$
correlation matrices built
from a variety of interpolators and smearing levels.
Using FLIC fermions to access the light quark mass regime, we explore
the curvature encountered in the energy of the states as the chiral
limit is approached.  We report a low-lying Roper state
contrasting earlier results using correlation matrices. To the best
of our knowledge, this is the first time a low-lying Roper resonance
has been found using correlation matrix methods. Finally, we present
our results in the context of the Roper results reported by
other groups.}

\FullConference{The XXVII International Symposium on Lattice Field Theory\\
                 July 26-31, 2009\\
                 Peking University, Beijing, China}

\begin{document}

\section{Introduction}

The first positive-parity excited state of the nucleon, known as the
Roper resonance, $N^{{\frac{1}{2}}^{+}}$(1440 MeV) ${\rm P}_{11}$, has
been a long-standing puzzle
since its discovery in the 1960's due to its lower mass compared to the adjacent
negative parity, $N^{{\frac{1}{2}}^{-}}$(1535 MeV) ${\rm S}_{11}$, state.    
In constituent quark models with harmonic oscillator
potentials, the lowest-lying odd parity state naturally occurs below
the ${\rm P}_{11}$ state (with principal quantum number $N=2$)
~\cite{Isgur:1977ef,Isgur:1978wd}, whereas, in nature the Roper
resonance is almost 100 MeV below the ${\rm S}_{11}$ state. 

Lattice QCD is very successful in computing many properties of hadrons
from first principles. In particular, in hadron spectroscopy, the
ground states of the hadron spectrum are now well understood. However, the
excited states still prove a significant challenge. The
first detailed analysis of the positive parity excitation of the nucleon
was performed in Ref.~\cite{Leinweber:1994nm} using Wilson fermions
and an operator product expansion spectral ansatz. Since then several
attempts have been made to address these issues in the lattice
framework, but in many cases no potential identification of the Roper state has
been made. Recently, however, in the analysis of
Refs.~\cite{Lee:2002gn,Mathur:2003zf,Sasaki:2005ap}
 a low-lying Roper state has been identified using Bayesian techniques.

Another state-of-the-art approach in hadron spectroscopy is the `variational
method' ~\cite{Michael:1985ne,Luscher:1990ck}, which is
based on a correlation matrix analysis. The identification of the
Roper state with this method wasn't successful in the past. However, very
recently, in Ref.~\cite{Mahbub:2009aa} a low-lying Roper
state has been identified with this approach employing a diverse
range of smeared-smeared correlation functions.

Here we discuss the new correlation matrix construction for isolating
the puzzling Roper state ~\cite{Mahbub:2009aa} and present our
Roper results in the context of the previous results reported by other
groups in recent times.

\section{Variational Method}
The two point correlation function matrix for $\vec{p} =0$ can be written as,
\begin{align}
G_{ij}(t) &= (\sum_{\vec x}{\rm Tr}_{\rm sp}\{ \Gamma_{\pm}\langle\Omega\vert\chi_{i}(x)\bar\chi_{j}(0)\vert\Omega\rangle\}), \\
          &=\sum_{\alpha}\lambda_{i}^{\alpha}\bar\lambda_{j}^{\alpha}e^{-m_{\alpha}t},
\end{align}
where, Dirac indices are implicit. Here, $\lambda_{i}^{\alpha}$ and
$\bar\lambda_{j}^{\alpha}$ are the couplings of interpolators $\chi_{i}$ and
$\bar\chi_{j}$ at the sink and source respectively and $\alpha$
enumerates the energy eigenstates with mass
$m_{\alpha}$. $\Gamma_{\pm}$ projects the parity of the eigenstates.\\ 
 Since the only $t$ dependence comes from the exponential term, one
 can seek a linear superposition of interpolators,
 ${\bar\chi}_{j}u_{j}^{\alpha}$, such that,  
\begin{align}
G_{ij}(t+\triangle t)\, u_{j}^{\alpha} & = e^{-m_{\alpha}\triangle
  t}\, G_{ij}(t)\, u_{j}^{\alpha},
\end{align}  
for sufficiently large $t$ and $t+\triangle t$. More detail can be found in Refs.~\cite{Melnitchouk:2002eg,Mahbub:2009nr,Blossier:2009kd}. Multiplying the above equation by $[G_{ij}(t)]^{-1}$ from the left leads to an eigenvalue equation,
\begin{align}
[(G(t))^{-1}G(t+\triangle t)]_{ij}\, u^{\alpha}_{j} & = c^{\alpha}\, u^{\alpha}_{i},
 \label{eq:right_evalue_eq}
\end{align} 
where $c^{\alpha}=e^{-m_{\alpha}\triangle t}$ is the eigenvalue. Similar to Eq.(\ref{eq:right_evalue_eq}), one can also solve the left eigenvalue equation to recover the $v^{\alpha}$ eigenvector,
\begin{align}
v^{\alpha}_{i}\, [G(t+\triangle t)(G(t))^{-1}]_{ij} & = c^{\alpha}v^{\alpha}_{j}.
\label{eq:left_evalue_eq}
\end{align} 
The vectors $u_{j}^{\alpha}$ and $v_{i}^{\alpha}$  diagonalize the correlation matrix at time $t$ and $t+\triangle t$ making the projected correlation matrix,
\begin{align}
v_{i}^{\alpha}G_{ij}(t)u_{j}^{\beta} & \propto \delta^{\alpha\beta}.
 \label{projected_cf} 
\end{align} 
The parity projected, eigenstate projected correlator, 
\begin{align}
 G^{\alpha}_{\pm}& \equiv v_{i}^{\alpha}G^{\pm}_{ij}(t)u_{j}^{\alpha} ,
 \label{projected_cf_final}
\end{align}
 is then analyzed using standard techniques to obtain masses of different states. 

\section{Lattice Details}
Our lattice ensemble consists of 200 quenched configurations with a lattice
volume of $16^{3}\times 32$. Gauge field configurations are generated
using the DBW2 gauge action
~\cite{Takaishi:1996xj,deForcrand:1999bi} and an
${\mathcal{O}}(a)$-improved FLIC fermion action ~\cite{Zanotti:2001yb} is
used to generate quark propagators.
The lattice spacing is $a=0.1273$ fm,
as determined by the static quark potential, with the scale set using
the Sommer scale, $r_{0}=0.49$ fm ~\cite{Sommer:1993ce}. In the
irrelevant operators of the fermion action we apply four sweeps of
stout-link smearing.
Various
sweeps (1, 3, 7, 12, 16, 26, 35, 48) 
 of gauge invariant Gaussian smearing
~\cite{Gusken:1989qx} are applied at the source (at $t=4$)
and at the sink.
The analysis is performed on ten different quark masses corresponding to pion masses $m_{\pi}=\{0.797,0.729,0.641,0.541,0.430,0.380,0.327,0.295,0.249,0.224\}$ GeV.
Error analysis is
performed using a second-order single elimination jackknife method,
where the ${\chi^{2}}/{\rm{dof}}$ is obtained via a covariance matrix
analysis method. Our fitting method is discussed extensively in Ref.~\cite{Mahbub:2009nr}.\\
The nucleon interpolator we consider is the local scalar-diquark
interpolator ~\cite{Leinweber:1994nm,Leinweber:1990dv},
\begin{align}
\chi_1(x) &= \epsilon^{abc}(u^{Ta}(x)\, C{\gamma_5}\, d^b(x))\,
u^{c}(x).
\label{eqn:chi1_interpolator}
\end{align}

\section{Results}
We consider several $4\times 4$ correlation matrices. Each matrix is constructed with different sets of correlation functions, each set element corresponding to
a different number of sweeps of gauge invariant Gaussian smearing at the
source and sink of the $\chi_{1}\bar\chi_{1}$ correlators. This provides a
large basis of operators with a wide range of overlaps among energy
states. 
 \begin{figure*}[!hpt] 
  \begin{center}
 \includegraphics [height=0.80\textwidth,angle=90]{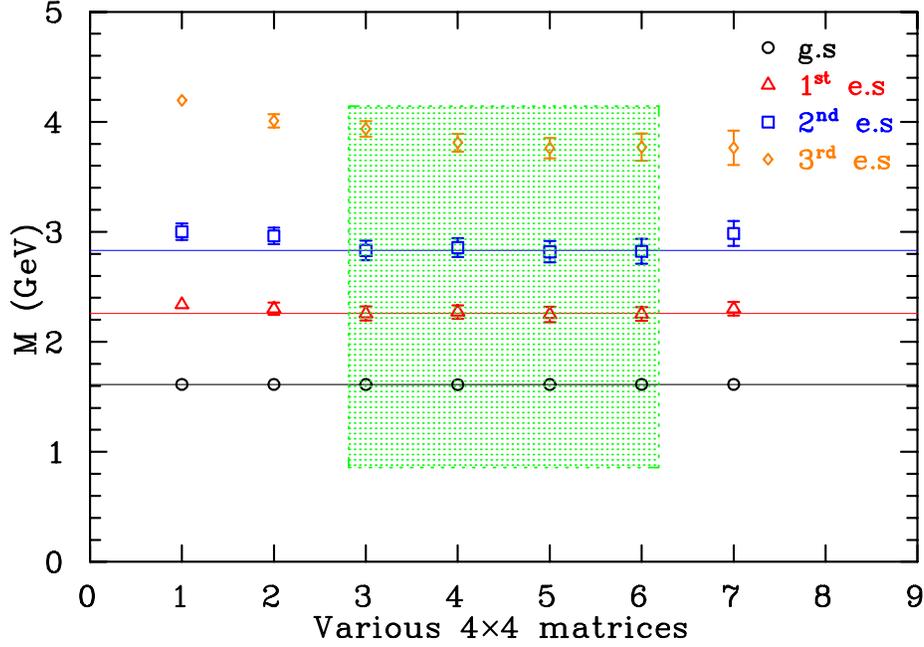}    
    \caption{(Color online). Masses of the nucleon, $N^{{\frac
          {1}{2}}^{+}}$ - states, from projected correlation functions
      as shown in Eq.(\protect\ref{projected_cf_final}) 
      for the pion mass of 797 MeV. Numbers in
      the horizontal scale correspond to each combination of smeared $4\times 4$
      correlation matrices. For instance, 1 and 2 correspond to the combinations
      of (1,7,16,35) and (3,7,16,35) respectively and so on, as
      discussed in the text following
      Eq.(\protect\ref{eqn:chi1_interpolator}). Masses are extracted according
      to the selection criteria described in 
      Ref.~\cite{Mahbub:2009nr}. The horizontal lines are drawn for the
      average mass over the four combinations (from $3^{\rm rd}$ to
      $6^{\rm th}$ as shown inside the shaded box).} 
   \label{fig:mass_for_all_combinations_Q1}
  \end{center}
\end{figure*}

We consider seven smearing combinations \{1=(1,7,16,35), 2=(3,7,16,35),
3=(1,12,26,48), \\ 4=(3,12,26,35), 5=(3,12,26,48),  6=(12,16,26,35),
7=(7,16,35,48)\} of $4\times 4$ matrices. 
In Ref.~\cite{Mahbub:2009nr} it was shown that one cannot isolate a
low-lying excited eigenstate using a single fixed-size source
smearing. The superposition of states manifested itself as a
smearing dependence of the effective mass.

In Fig.\ref{fig:mass_for_all_combinations_Q1}, masses extracted from
all the combinations of $4\times 4$ matrices (from $1^{\rm st}$ to $7^{\rm
  th}$) are shown for the pion mass of 797 MeV. Some dependence of
the excited states on smearing sweep count is observed here as in
Ref.~\cite{Mahbub:2009nr} for a few of the interpolator basis smearing
sets. However the ground and first excited states are robust against
changes in the interpolator basis, providing evidence that an energy
eigenstate has been isolated. It should be noted
that the highest excited state (the third excited state) is influenced more
by the level of smearing than the lower excited states. This is to be
expected as this state must accommodate all remaining spectral
strength and this is dependent on smearing.    

\begin{figure*}[!hpt] 
  \begin{center}
 \includegraphics [height=0.80\textwidth,angle=90]{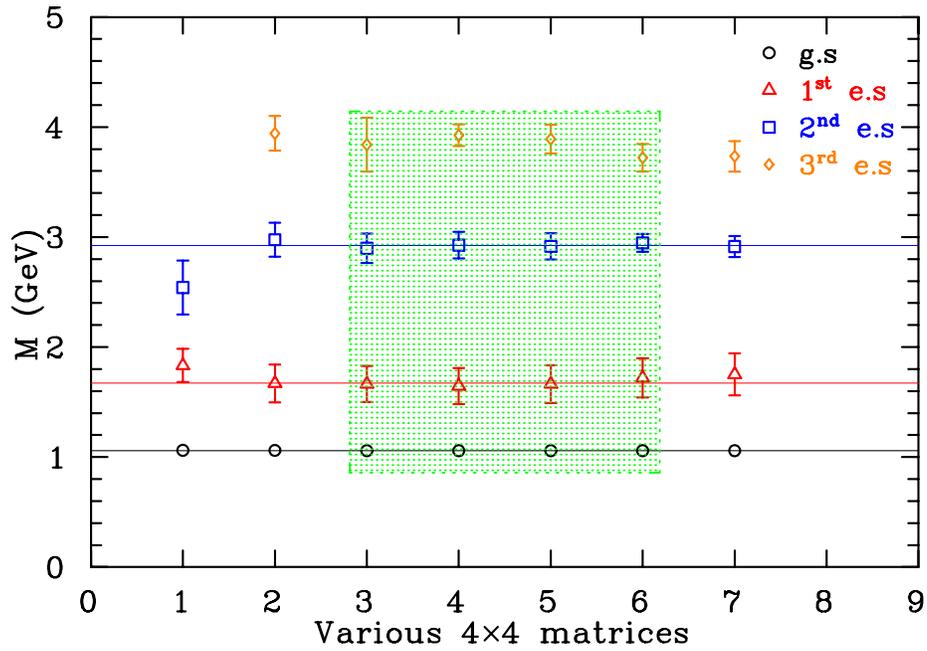}    
    \caption{(Color online). As in
      Fig.\protect\ref{fig:mass_for_all_combinations_Q1}, but for the light pion
      mass of 249 MeV.} 
   \label{fig:mass_for_all_combinations_Q6}
  \end{center}
\end{figure*}

The $1^{\rm{st}}$ combination in
Fig.\ref{fig:mass_for_all_combinations_Q1} provides heavier excited
states as this basis begins with a low number of smearing sweeps (a
sweep count of 1) and also contains another low smearing set of 7
sweeps. The first and second excited states sit a little high in
comparison with  the other
bases. This is also evident in
Fig.\ref{fig:mass_for_all_combinations_Q6} for the lighter quark mass.
 Hence, extracting masses with this basis is not as reliable as
other sets containing a great diversity of large numbers of smearing sweeps. The
$2^{\rm{nd}}$ combination also contains elements with a small smearing
sweep count (3 and 7), hence this basis also provides heavier excited
states and shows some systematic drift in the second excited
state. However, this basis has reduced contamination from the excited
states when compared with the first basis.

\begin{figure*}[!hpt] 
  \begin{center} 
 \includegraphics [height=0.80\textwidth,angle=90]{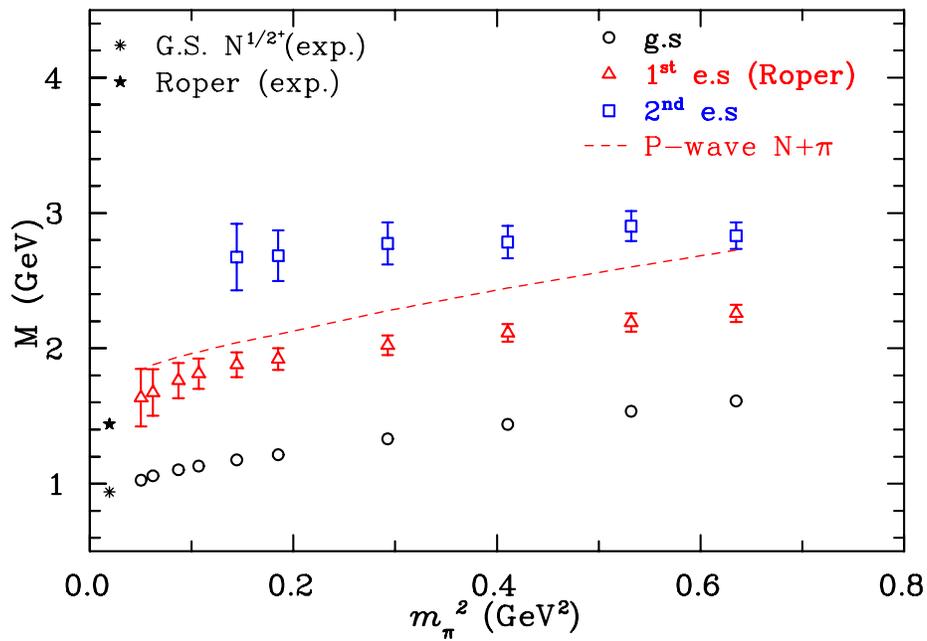}    
    \caption{(Color online). The ground, the Roper and second excited states
      and the non-interacting
      P-wave $N+\pi$ are illustrated. The black filled symbols are the
      experimental values of the ground and the Roper states.}
   \label{fig:paper_m.avg_4com.x1x1.4x4.sqrt_erravg_4combs_errbasis_4combs.2states.allQ.NEW}
  \end{center}
\end{figure*}

We can observe at this point that including basis elements with 2
consecutive low
smearing sweep counts (for
instance, consecutive low numbers of smearing sweeps 1,7 of
$1^{\rm{st}}$ combination and 3,7 of $2^{\rm{nd}}$ combination,
respectively), provides a basis which does not span the space
well.  We also observe that
the inclusion of basis elements with 2 consecutive high levels of smearing (for
instance, a sweep count of 35, 48 as in the $7^{\rm th}$ combination)
does not span the space well and gives rise to larger uncertainties.

\begin{figure*}[!hpt] 
  \begin{center} 
 \includegraphics [height=0.98\textwidth,angle=90]{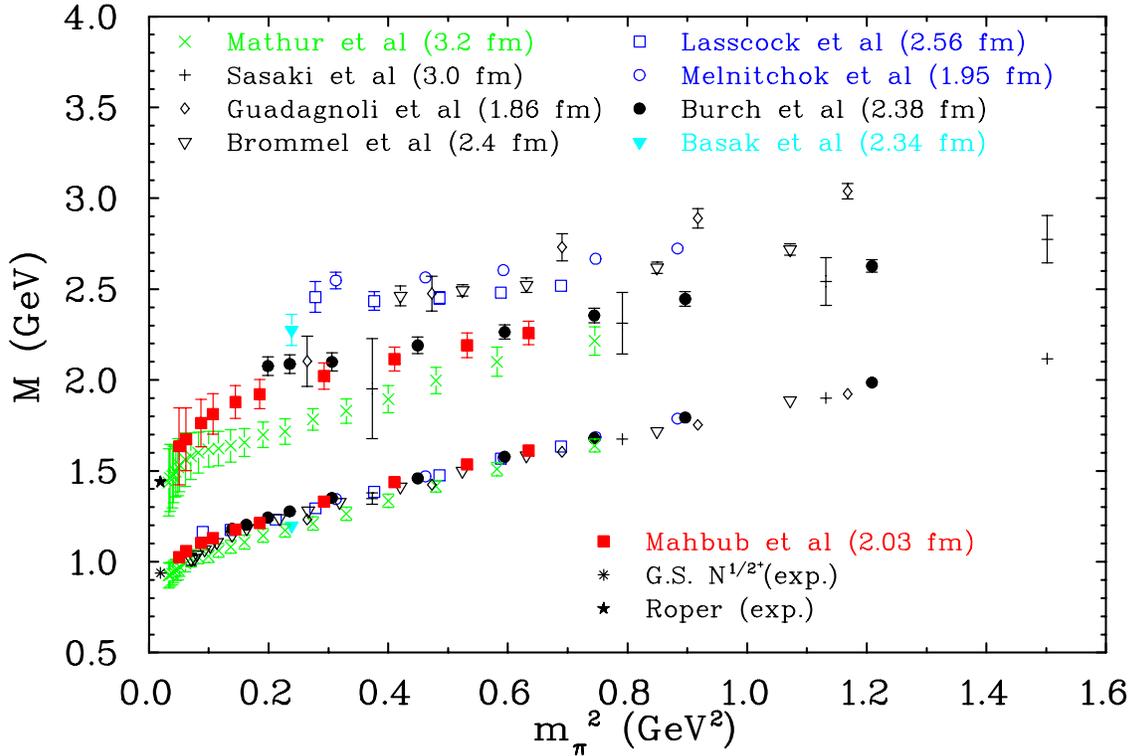}    
    \caption{(Color online). Compilation of current lattice results of
      the nucleon, $N^{{\frac{1}{2}}^{+}}$, for the ground and first
      excited states reported in
      Refs.~\cite{Mathur:2003zf,Sasaki:2005ap,Mahbub:2009aa,Melnitchouk:2002eg,Guadagnoli:2004wm,Brommel:2003jm,Lasscock:2007ce,Burch:2006cc,Basak:2006ww}
      as described in the text.}
   \label{fig:nucleon.mass.spectrum.world.NEW}
  \end{center}
\end{figure*}

The $3^{\rm rd}$, $4^{\rm th}$ and $5^{\rm th}$ combinations are well spread over
the given range of smearing sweeps.
They don't include successive lower smearing sweep counts. The $5^{\rm
  th}$ combination contains the basis element with a sweep count of 48
but has only slightly larger statistical errors than the $4^{\rm th}$
basis choice. All these bases provide diversity.
 It is observed that the $3^{\rm rd}$ through the $6^{\rm th}$ combinations
provide consistent results for the first and second excited states. 
An analysis is performed to calculate the systematic errors associated
with the choice of basis over the preferred four combinations (from
$3^{\rm rd}$ to $6^{\rm th}$) with
$\sigma_{b}=\sqrt{{\frac{1}{N_{b}-1}}\sum_{i=1}^{N_{b}}(M_{i}-\bar{M})^{2}}$,
where, $N_{b}$ is the number of bases, in this case equal to 4. The
statistical and systematic errors due to basis choices are added in
quadrature, $\sigma=\sqrt{\bar{\sigma}_{s}^{2}+\sigma_{b}^{2}}$, which
is shown in 
Figs.~\ref{fig:paper_m.avg_4com.x1x1.4x4.sqrt_erravg_4combs_errbasis_4combs.2states.allQ.NEW}
and \ref{fig:nucleon.mass.spectrum.world.NEW}.

In
Fig.~\ref{fig:paper_m.avg_4com.x1x1.4x4.sqrt_erravg_4combs_errbasis_4combs.2states.allQ.NEW},
it is interesting to note that the observed lattice Roper state sits
lower than the P-wave $N+\pi$, indicative of attractive $\pi {\rm N}$
interactions producing a resonance at physical quark masses. 

In Fig.~\ref{fig:nucleon.mass.spectrum.world.NEW}, the ground state
results are consistent for all the reported works. However,
significant differences are readily observed in the results of the
first positive parity excited (Roper) state. In
Ref.~\cite{Mathur:2003zf}, Mathur ${\it et \, al.}$ used a constrained
curve fitting method. In Ref.~\cite{Sasaki:2005ap}, Sasaki $\it{et \,
  al}.$ used the Maximum Entropy Method. In
Ref.~\cite{Guadagnoli:2004wm}, Guadagnoli $\it{et \, al}.$ used a
modified  correlator  technique. Refs.~\cite{Mahbub:2009aa,Melnitchouk:2002eg, Brommel:2003jm, Lasscock:2007ce, Burch:2006cc, Basak:2006ww} all used the variational
approach. Brommel $\it{et \, al}.$ ~\cite{Brommel:2003jm} considered 
standard nucleon interpolators $\chi_{1} \, ,\chi_{2}$ and $\chi_{3}$ and 
employ Jacobi smeared sources. They have performed simulations in a range of
pion masses of 270-866 MeV and used $3\times 3$ correlation
matrices. They reported results that are too high to be interpreted as
the Roper  state. 
The early analysis of CSSM Collaboration by Melnitchouk $\it{et \, al}.$
~\cite{Melnitchouk:2002eg} used 
$\chi_{1}$ and $\chi_{2}$ interpolators and used a single Gaussian source
smearing of 20 sweeps. They found no
evidence for the Roper state as the energy states of their $2\times 2$
correlation matrix analysis also proved too high to be interpreted as the Roper resonance. The results from  
Lasscock $\it{et \, al}.$ ~\cite{Lasscock:2007ce} from $3\times 3$  
correlation matrix analysis of standard interpolators also sits 
 as high as those of Brommel and Melnitchouk, and cannot be
interpreted  as the
Roper state. Burch $\it{et \, al}.$ ~\cite{Burch:2006cc}
considered the alternative approach of using Jacobi smeared sources of two
different widths (narrow and wide) to increase the basis of operators
and  performed $6\times6$ correlation matrix 
analysis for a pion mass down to 450 MeV. Though their lightest quark
mass results sit slightly above our results labeled Mahbub $\it{et \,
  al}.$ in Fig.~\ref{fig:nucleon.mass.spectrum.world.NEW}
~\cite{Mahbub:2009aa}, in the
heavy quark-mass region the significant overlap between the results of
Burch $\it{et \, al}.$ and ours is apparent. Basak $\it{et \, al}.$
~\cite{Basak:2006ww} used non-local operators to 
form the basis of their correlation matrix and simulated at a pion mass
of 490 MeV. Their results are also high. They reported that they did
not find any Roper like positive-parity excitation. Using  
smeared-smeared correlators to construct a basis of correlation matrices,
 we identified a low-lying Roper state with significant curvature
 apparent at  lighter quark masses ~\cite{Mahbub:2009aa}.

\section{Conclusions}
Through the use of a variety of smeared-smeared
correlation functions in constructing 
correlation matrices, the first positive parity excited state of the nucleon
$N^{{\frac{1}{2}}^{+}}$, the Roper state, has been observed for the
first time using the variational analysis ~\cite{Mahbub:2009aa}. The
current status of results for the  
Roper resonance from a number of groups are reviewed. Our lattice
Roper  state has  a
tendency to approach the physical Roper state showing a significant
curvature as the chiral limit is approached
~\cite{Mahbub:2009aa}. This work signifies the importance of using a
diverse range of smeared-smeared correlation functions when
constructing correlation matrices for the identification of the elusive Roper
state.  This robust approach should also be applied for larger
dimensions of the correlation matrices not only built from the
$\chi_{1}\bar\chi_{1}$ correlators but also using
$\chi_{1}\bar\chi_{2}$ and in the negative parity channel. This will
be the subject of future investigations.

\end{document}